\documentclass[aps,prl,twocolumn,showpacs,groupedaddress]{revtex4}

\usepackage{graphicx}
\usepackage{dcolumn}
\usepackage{amsmath}

\begin{document}

\title{Signatures of interband scattering in spectroscopic experiments on $MgB_2$}
\author{C. P. Moca and B. Jank\' o}
\affiliation{Material Science Division, Argonne National
Laboratory, Argonne, Illinois, 60439\\ Department of Physics,
University of Notre Dame, Notre Dame, Indiana, 46556}

\begin{abstract}
Within a two-band, strong coupling model we analyze
$SIS$-tunneling conductance, Raman scattering and optical
conductivity measured on $MgB_2$ samples. We find that features
observed in tunneling and Raman spectroscopy at intermediate
energies $\omega=\Delta_{\sigma}+\Delta_{\pi}\simeq 8\div 10meV$
can be consistently explained when incoherent scattering between
$\sigma$ and $\pi$ band is considered. We calculate the optical
conductivity and find that strong coupling effects are important
for explaining the presently available data, and predict the
location and magnitude of interband scattering features for this
spectroscopic probe as well.

\end{abstract}

\maketitle

\emph{ Introduction.} The discovery of superconductivity in
$MgB_2$ with a critical temperature $T_C=39K$ \cite{Nagamatsu} has
initiated intense research efforts on elucidating the properties
of this surprisingly simple compound. The main reason behind the
relatively high transition temperature is still unknown. However,
there is a growing consensus that strong electron-phonon
interaction \cite{Kortus} is responsible for superconductivity in
$MgB_2$. The large isotope effect measured by Bud'ko {\it et al.}
\cite{Budko} for boron ($\alpha=0.26\pm 0.03$) clearly shows that
phonons associated with boron vibrations play a significant role
in superconductivity in this compound.  Tunneling and point
contact spectroscopy experiments \cite{Karapetrov, Rubio, Szabo,
Sharoni, Kohen, Schmidt} reveal a distribution of energy gap
$2\Delta(0)/T_C$ between 1.1 and 4.5. Recent experiments of
Giubileo {\it et al.} \cite{Giubileo} pointed the presence of two
gaps. Evidence of two gaps were found also in the specific heat
\cite{Bouquet} and recently in Raman scattering experiments
\cite{Chen}.

According to Liu {\it et al.} \cite{Liu} the four Fermi surface
sheets in $MgB_2$ can be grouped into quasi two-dimensional
$\sigma$ bands and three dimensional $\pi$ bands, and the normal
and superconducting properties of $MgB_2$ can be described by an
effective two-band model. Within this model they calculated the
coupling constants and energy gap ratios in the weak coupling
regime. Later Golubov {\it et al.}\cite{Golubov} provided the
two-band decomposition of the superconducting Eliashberg functions
$\alpha^2F_{ij}(\omega)$ (where $i$ and $j$ denote $\sigma$ and
$\pi$ bands), which describe the electron-phonon coupling in
$MgB_2$ as function of frequency. While electron-phonon
interaction was clearly investigated in large spectrum of
approximations, the role of scattering between the bands was not
analyzed in detail.

We turn our attention to the theory of spectroscopic properties of
$MgB_2$ because, similarly to the situation in cuprate
superconductors, the most severe constrains on theoretical models
are imposed by spectroscopic measurements. In fact, the aim of
this paper is to point out that a {\em simultaneous} explanation
of the main features in $SIS$ tunnel spectroscopy\cite{Schmidt}
and Raman scattering \cite{Chen} can only be done if one allows
for the existence of incoherent (momentum non-conserving)
single-particle scattering events between. We have also analyzed
recent optical conductivity results \cite{Kaidl}. We find, as
shown in detail below, that agreement between theory and
experiment is substantially improved over the weak-coupling
analysis by considering strong coupling effects alone. We have
also calculated the effect of incoherent interband scattering on
such measurements. While the presently available data does not
allow us to firmly conclude, that there is evidence for incoherent
interband scattering in the optical conductivity data as well, it
does not contradict the presence of the process either. We
nevertheless present our predictions for optical conductivity in
the presence of interband scattering, since we believe that our
predictions for this probe will be useful in analyzing future,
more accurate measurements.

\emph{ Theoretical Model.} We will adopt a two-band Eliashberg
scheme to calculate the various spectroscopic quantities. The weak
coupling version of the model for two-band superconductivity was
originally proposed by Suhl, Matthias and Walker\cite{Suhl}.
However, due to the lack of a genuine two-band superconductor at
that time, interest in this model gradually decreased. More
recently, the Suhl-Matthias-Walker model was revived and applied
to cuprates by Kresin\cite{Kresin}. We have used a similar
model\cite{Moca} to calculate the penetration depth in $MgB_2$,
and we are now applying this framework to derive the spectroscopic
properties as well. The starting point of our analysis is the
two-band Hamiltonian ${\cal H} ={\cal H}_0+{\cal H}_I+{\cal
H}_{\rm i-b}$. Here ${\cal H}_0$ is describing the free electrons,
\begin{equation}
{\cal H}_0=\sum\limits_{\mathbf{k},\sigma }\varepsilon
_{\mathbf{k}}c_{\mathbf{k} \sigma }^{+}c_{\mathbf{k}\sigma
}+\sum\limits_{\mathbf{k},\sigma }\xi _{
\mathbf{k}}d_{\mathbf{k}\sigma }^{+}d_{\mathbf{k}\sigma
}+\sum\limits_{ \mathbf{q}}\omega
_{\mathbf{q}}b_{\mathbf{q}}^{+}b_{\mathbf{q}}. \label{h0}
\end{equation}
In the above expression $c_{ \mathbf{k}\sigma }^{+}$ $\left(
c_{\mathbf{k}\sigma }\right) $ creates (annihilates) electrons in
$\sigma$ band and $d_{\mathbf{k}\sigma }^{+}$ $ \left(
d_{\mathbf{k}\sigma }\right) $ are similar operators acting in
$\pi$ band. The last term describes the non-interacting phonons
with the dispersion given by $\omega _{\mathbf{q}}$. The second
term in the Hamiltonian, ${\cal H}_{I}$ corresponds to the
interaction of the conduction electrons with the phonons,
\begin{eqnarray}
{\cal H}_I &=&\sum\limits_{\mathbf{k,q},\sigma }V_{\sigma
\sigma}(\mathbf{k},\mathbf{q})c_{ \mathbf{k+q}\sigma
}^{+}c_{\mathbf{k}\sigma }(b_{\mathbf{q}}+b_{-\mathbf{q} }^{+})
\nonumber \\ &&+\sum\limits_{\mathbf{k,q},\sigma
}V_{\pi\pi}(\mathbf{k},\mathbf{q})d_{\mathbf{\ k+q }\sigma
}^{+}d_{\mathbf{k}\sigma }(b_{\mathbf{q}}+b_{-\mathbf{q}}^{+})
\label{3} \\ &&+\sum\limits_{\mathbf{k,q},\sigma }V_{\sigma
\pi}(\mathbf{k},\mathbf{q})\left[ c_{ \mathbf{k+q}\sigma
}^{+}d_{\mathbf{k}\sigma }(b_{\mathbf{q}}+b_{-\mathbf{q}
}^{+})+h.c.\right]. \label{hi}
\end{eqnarray}
Finally we introduce a new term ${\cal H}_{\rm i-b}$ to account
for the \emph{ incoherent (momentum non-conserving) interband
scattering}
\begin{equation}
{\cal H}_{\rm i-b}=\sum\limits_{\mathbf{k},\sigma
}(T_{\mathbf{kq}} c_{\mathbf{k} \sigma }^{+}d_{\mathbf{q}\sigma} +
h.c.) \label{hib}
\end{equation}.
The effect of this last term is the main focus of our study, and
we will show below, that it plays an important role in explaining
the observed features in spectroscopic measurements.

Our approach is based on the self-consistent solution of the
Eliashberg equations corresponding to the Hamiltonian
(\ref{h0}-\ref{hib}). Self-consistency is reached on the imaginary
axis, and Pad\'e approximation \cite{Vidberg} is then used to
analytically continue to the real axis. This procedure gives
accurately the frequency dependence of the gap functions
$\Delta_{\sigma}(\omega)$ and $\Delta_{\pi}(\omega)$ and the
renormalization factors $Z_{\sigma}(\omega)$ and $Z_{\pi}(\omega)
$. The normal self-energy contribution to the $i$ band due to the
direct scattering from and in the $j$ band can be written as $
1/(2\tau_{ij})g_j (i\omega _n)$ where $g_i(i\omega _n)=\omega
_n/{\sqrt{\omega _n^2+\Delta _i^2(\omega _n)} }$ and the anomalous
self-energy can be written as $ 1/(2\tau_{ij})f_j (i\omega _n)$
where $f_i(i\omega _n)=\Delta _i/{\sqrt{\omega _n^2+\Delta
_i^2(\omega _n)} }$.  The structure in $\Delta (\omega )$ usually
gives rise to observable variations in the density of states at a
frequency comparable with the sum of phonon energy and gap value.
However, in $MgB_2$ the phonon mode most relevant to
superconductivity is $E_{2g}\simeq 67meV$, and observation of
density of states features at such high energies are difficult to
be observed experimentally. For the present calculation we model
the phonon density with a lorentzian: $ F(\omega) = A/((\omega
-\omega _1)^2+\omega _2^2)- A/(\omega _3^2+\omega _2^2)$  for
$\left| \omega -\omega _1\right| <\omega _3 $ and zero otherwise.
Here the frequency parameters are: $\omega _1=67meV$, $\omega
_2=3.5meV$, and $\omega _3=30meV$. With these parameters, and
taking a cut-off  frequency at $\omega _c=200meV$, we fix $A=1.31$
by imposing $\int\limits_0^{\omega _c}F(\omega )d\omega =1$.  The
best fit to the experimental temperature dependence of the gap
function\cite{Buzea} and with the observed $T_C=39K$  was found
for $\alpha _{\sigma \sigma}^2=17.5$, $\alpha _{\pi \pi}^2=7.0$,
$\alpha _{\sigma \pi}^2=1.0$, $\alpha _{\pi \sigma}^2=2.0$. The
effect of Coulomb pseudopotential have been considered through the
renormalized parameters $\mu _{ii}^{\star }=0.1$ and $ \mu
_{ij}^{\star }=0$ for $i\neq j$. The effect of cross-band
interaction can be analyzed in terms of $\Gamma _{ij}=
\tau^{-1}_{ij}$ which directly describe the scattering rate.  This
is similar with solving the Eliashberg equation for an alloy where
similar terms appear due to scattering on impurities
\cite{Carbotte}. Throughout of our calculations we have considered
that $\Gamma_{ii}=0$ (no intra-band scattering) and that only
$\Gamma_{ij}\neq 0$ for $i\neq i$.

\emph{ Break Junction Tunneling.} We start the comparison to
experiment by analyzing the break junction $SIS$ tunneling
experimental data of Schmidt {\it et al.} \cite{Schmidt}. A
careful analysis of the data reveals the presence a dip at a
frequency $\omega \sim \Delta_{\sigma}+\Delta_{\pi}\sim 10 meV $
and no measurable effects are seen near the frequency
corresponding to the larger gap. This suggests that, due to its
$3D$ nature,  only the $\pi$ band gives a contribution to the
tunneling current in break junctions, while the $\sigma$ band, due
to its quasi $2D$ nature, has little overall contribution. The
electron-phonon interaction alone does not give a satisfactory
explanation for the above mentioned feature, and a new term is
needed.  This term describes incoherent single particle tunneling
between bands, and is formally similar to the proximity term first
proposed by McMillan \cite{McMillan}. Within this extended
Suhl-Matthias-Walker model, the superconductivity in $\pi$ band
can be thought of as being induced by the $\sigma$-band.
Scattering between bands diminishes the larger gap while the
smaller one is enhanced. More importantly, however, self-energy
contributions in lowest order in interband scattering will give
rise to intermediate states of energies $\Delta_{\sigma} +
\Delta_{\pi } \sim 8-10 \, {\rm meV} $, precisely at the voltage
where the dip in the tunnel conductance is seen.

\begin{figure}[tbp]
\centerline{\includegraphics[width=3.5in]{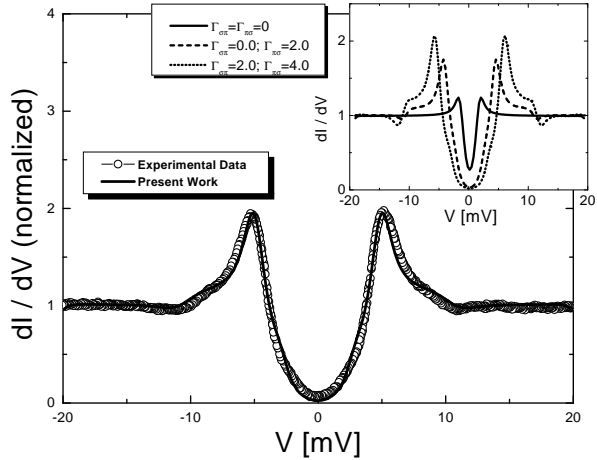}}
\vspace*{3ex} \caption{  Results for normalized $\pi$-band $SIS$
tunneling conductance (solid line) compared with the experimental
data of Schmidt {\it{ et al. }}\cite{Schmidt} (open circle
symbols). The scattering rates used in this calculations are
$\Gamma_{\pi,\sigma}=2.5$ and $\Gamma_{\sigma,\pi}=0.5$.
 The inset shows the calculated $SIS$ tunneling
conductance for different scattering rates.} \label{SISTunneling}
\end{figure}

In order to study the effect of the direct inter-band interaction
we have first calculated the density of states for each band by
using the well known relation $ N_S^{(i)}(\omega
)=N_N^{(i)}(\omega )Re\left( \left( \omega-\Gamma_T\right )/
\sqrt{(\omega-\Gamma_T)^2-\Delta _i^2(\omega )}\right)$,  where
$N_S(\omega )\,$ and $N_N(\omega )$ are the energy dependent
density of states of the material in the superconducting and
normal state. We have calculated first the  superconducting
density of states and then using these results we investigate the
$SIS$ conductance by calculating first the tunneling current by a
direct convolution:
\begin{equation}
I_{i,j}(eV)\sim \int\limits_{-\infty }^{+\infty }\left[ n_F(\omega
)-n_F(\omega +eV)\right] N_S^{(i)}(\omega )N_S^{(j)}(\omega +eV)
\label{Conductance}
\end{equation}
and then taking the derivative as function of bias. In Eq.
(\ref{Conductance}) $n_F(\omega)$ is the Fermi-Dirac function and
$N_S^{(i)} (\omega)$ with $i=\pi, \sigma$ is the density of states
corresponding to one of the bands. The conductance can be found by
differentiating the tunneling current as function of bias and a
direct quantitative comparison with the experimental data is
possible. Fig. \ref{SISTunneling} represents the results obtained
for the conductance for the $\pi$-band together with the
experimental data of Schmidt {\it et al.} \cite{Schmidt}. The
interband self-energy leads to modification in the structure of
the gap function at energies
$\omega=\Delta_{\sigma}+\Delta_{\pi}$. Increasing the scattering
rates between the bands lead to $(i)$ an enhance of the smaller
gap corresponding to the $\pi$ band and to a decreasing of the
$\sigma$-band gap; $(ii)$ increasing of the dip feature at the
energy $\Delta_{\sigma}+\Delta_{\pi}$ in the $SIS$ conductance
spectrum. The last effect can be verified experimentally by
comparing samples of different impurity concentrations. The
presence of impurities will increase such interband scattering
processes, and an interesting and somewhat counter-intuitive
increase in gap size is therefore expected with the increase of
impurity concentration. This would be a qualitative consistency
check on our theory.

\emph { Raman Spectroscopy.} This is the second spectroscopic
experiment that we analyze in the same theoretical framework and
find an excellent agreement between the experiment and our model.
Early Raman experiments performed on $MgB_2$ were somewhat
inconsistent due to differences in sample quality: measurements on
poly-crystaline samples\cite{Chen} reveal the presence of two gaps
while single crystal measurements restricted to only $ab$-plane
polarized spectra show only the presence of a one superconducting
gap \cite{Quilty}. Recently a more careful analysis resolved this
inconsistency and the agreement is that the Raman spectrum does
reveal the presence of two bands\cite{Quilty1}.
\begin{figure}[tbp]
\centerline{\includegraphics[width=3.5in]{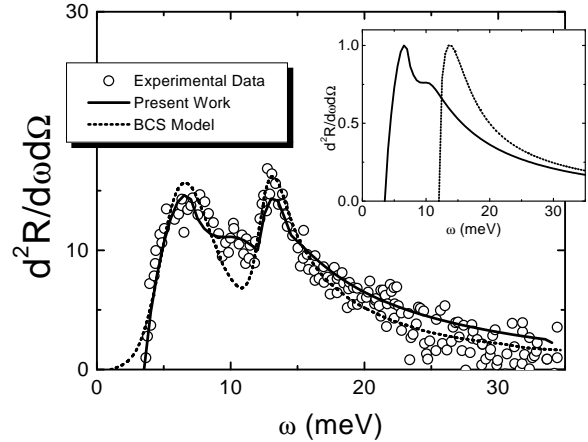}}
\vspace*{3ex} \caption{Calculated Raman spectra at temperature
$T=15K$ (solid line) compared with the experimental data of Chen
{\it et al.} \cite{Chen} (open symbols). The dotted line
represents the two-band BCS like model fit. The scattering rates
in these calculations were $\Gamma_{\sigma,\pi}=0.5$ and
$\Gamma_{\pi,\sigma}=3.0$. The inset presents the normalized
results obtained for the Raman spectrum for each band separately.}
 \label{RamanSpectrum}
\end{figure}
We have considered the effect of scattering between the bands when
calculating the Raman susceptibility \cite{Jiang}
$X(\mathbf{q}\rightarrow 0, \omega )$. The Raman spectrum was
calculated separately for each band and then the spectrum was
calculated as a weighted sum of contributions from each band. The
inset of Fig.\ref{RamanSpectrum} presents the normalized
contributions coming from each band. There is a complete depletion
of the Raman spectrum below  $2\Delta_i$ for each band and the
spectrum has a vertical slope at $2\Delta_i$. The weighting
coefficients are 14.61 for $\pi$-band and 4.39 for the
$\sigma$-band leading to a ratio of 3.32 for the contributions of
each band (in contrast with the 2.97 ratio obtained in the
Suhl-Matthias-Walker model by Chen {\it et al.}\cite{Chen}. In our
calculations the Raman vertex  was calculated for simplicity in
the limit $\gamma(\mathbf{p})=1$. Even in this limit the effect of
scattering between different bands is important, and leads to
significant modifications of the spectrum. Our results for the
Raman spectrum are presented in Fig.\ref{RamanSpectrum} and
compared to  the experimental data of Chen {\it et al.}
\cite{Chen}. The dashed line is the weak coupling $BCS$ two-band
model fit proposed by the same authors. The significant difference
between our results and theirs is the presence of the hump at an
energy $\omega=\Delta_{\sigma}+\Delta_{\pi}$, a feature that
theoretically can be obtained by considering only the scattering
between the bands, and does not appear in the simple
Suhl-Matthias-Walker model. In our calculations the best fit of
the experimental data was obtained when $\Gamma_{\sigma, \pi}=0.5$
and $\Gamma_{\pi, \sigma}=3.0$. The other parameters used in the
calculations are r $\alpha _{\sigma \sigma}^2=17.5$, $\alpha _{\pi
\pi}^2=5.5$, $\alpha _{\sigma \pi}^2=1.0$, $\alpha _{\pi
\sigma}^2=2.0$ and the same values for the Coulomb pseudopotential
as in the case of analysis of the tunneling experiment.

\begin{figure}[tbp]
\centerline{\includegraphics[width=3.5in]{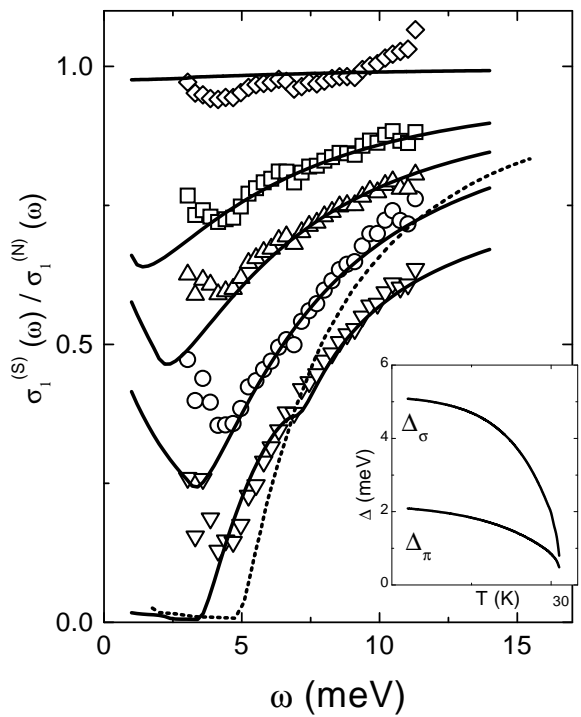}}
\vspace*{3ex} \caption{ Real part of conductivity ration
$\sigma_1^{(S)}(\omega)/\sigma_1^{(N)}(\omega)$  for different
temperatures (solid lines) from bottom to top $T=6K, 17.5K, 24 K,
27K, 30K$ compared with the experimental data extracted from Ref.
[\onlinecite{Kaidl}] (dotted symbols for the same temperatures
from bottom to top). The dotted line is the constant gap model for
temperature T=6K. The scattering rates are $\Gamma_{\sigma,
\pi}=0.5$ and $\Gamma_{\pi,\sigma}=2.5$. The inset presents the
temperature dependence of the gap functions for each band
separately.  } \label{OpticalConductivityRealPart}
\end{figure}

\emph {Optical Conductivity.} The last spectroscopic experiment
that we focus on is the optical conductivity, where the importance
of the scattering between the bands was analyzed in the framework
of Mattis-Bardeen theory \cite{Mattis}. A  quantitative comparison
with the experimental data on optical conductivity was possible in
this case as well, as shown in Fig.
\ref{OpticalConductivityRealPart}. One immediate conclusion is
already apparent: the overall fit to the experimental data is much
better in the Eliashberg theory than in the constant gap model.
Due to the small critical temperature obtained for the measured
sample \cite{Kaidl} $(T_C=30K)$ we had to modify the phonon
coupling constants in order to obtained the desired critical
temperature. Therefore we consider $\alpha _{\sigma
\sigma}^2=15.5$, $\alpha _{\pi \pi}^2=5.0$, $\alpha _{\sigma
\pi}^2=1.0$, $\alpha _{\pi \sigma}^2=2.0$, and for the scattering
rates the best fit with the experimental data was obtained for
$\Gamma_{\sigma, \pi}=0.5$ and $\Gamma_{\pi,\sigma}=2.5$. The
small feature appearing in the real part of the optical
conductivity at $\omega=\Delta_{\sigma}+\Delta_{\pi}\simeq 8meV$
(in this case this energy is smaller due to the smaller critical
temperature $\Delta_{\pi}(T=0K)=2.51 \, {\rm meV}$,
$\Delta_{\sigma}(T=0K)=5.48 \, {\rm meV}$), is due to the
modifications in the frequency dependence of the gap functions
induced by the effect of inter-band scattering in the electronic
self-energy. The small feature in the data near this frequency
appears to be an extrinsic effect \cite{Carnahan}, and is unlikely
to correspond to interband scattering. However, we believe that
future experiments will be able to detect the interband structure
we predict here.

\emph{ Conclusions.} We have investigated the role of the
incoherent inter-band scattering in the framework of two-band
strong coupling model. We found results consistent with three
experimental measurements. The main results of our work can
therefore be summarized as follows: $(i)$ superconductivity in
$MgB_2$ can be well explained in the framework of the two-band
model; $(ii)$ The features observed in tunneling spectroscopy, and
Raman scattering  at energies
$\omega=\Delta_{\sigma}+\Delta_{\pi}\simeq 8\div 10meV$ can be
consistently explained by considering the incoherent scattering
between the bands. While there is no clear feature observed in
optical conductivity at this frequency, we predict that such
features will eventually be observed, $(iii)$ Overall, the fit to
the experimental data  is much better if a strong coupling theory
is used, rather than a two-band, weak coupling, constant gap model.
\cite{Schmidt}.

\emph{ Acknowledgments.} We are especially indebted to Drs. H.
Schmidt and M. Carnahan for supplying us with tunneling and
optical spectroscopy data, respectively. We wish to thank them as
well as Drs. G. Arnold, G. W. Crabtree, K. Gray, M. Iavarone,
W.-K. Kwok, G. Karapetrov,  A. Koshelev, J. Orenstein, and J.
Zasadzinski for useful discussions.

\end{document}